\documentclass[12pt,preprint]{aastex}

\begin{document}

\title{
A massive warm baryonic halo in the Coma cluster}

\author{Massimiliano~Bonamente$\,^{1,2}$, Marshall~K.~Joy$\,^{2}$ and
Richard Lieu$\,^{1}$
}

\affil{\(^{\scriptstyle 1} \){Department of Physics, University of Alabama,
Huntsville, AL}\\
\(^{\scriptstyle 2} \){National Space Science and Technology Center, NASA/MSFC, Huntsville, AL}\\
}

\begin{abstract}
Several  deep PSPC observations of the Coma cluster reveal a  very large-scale halo 
of soft X-ray emission, 
substantially in excess of the well known radiation from the
hot intra-cluster medium.
The excess emission, previously reported in the central region of the cluster using 
lower-sensitivity EUVE and ROSAT data, is
now evident out to a radius of 2.6  Mpc, 
demonstrating that the soft excess radiation
from clusters is a phenomenon  of cosmological significance.
The X-ray spectrum at these large radii cannot be modeled non-thermally, but is consistent
with the original scenario of thermal emission from warm gas at $\sim$ $10^6$ K. The mass of
the warm gas is on par with that of the hot X-ray emitting plasma,  
and significantly more massive if the warm gas resides in low-density
filamentary structures. Thus the data lend vital support to current
theories of cosmic evolution, which predict that at low redshift $\sim$ 30-40 \% of the baryons reside in 
warm filaments converging at clusters of galaxies.
\end{abstract}

\section{Introduction}

Clusters of galaxies are strong emitters of X-rays, which originate from a hot and
diffuse intra-cluster medium (ICM). At the typical temperatures of a few $\times 10^{7}$ K,
the bulk of the hot ICM emission is detected at energies $\geq$ 1 keV, intervening
Galactic absorption being responsible for a substantial reduction of  flux below $\sim$ 1 keV.
However, hot gas is not the only high energy component seen in clusters, and
the extreme-ultraviolet (EUV) and soft X-ray band below $\sim$ 1 keV offer a unique window to investigate
the presence of other phases in the ICM.

In 1996, Lieu et al. (1996) reported the discovery of excess EUV and soft X-ray emission
above the contribution from the hot ICM in the Coma cluster; their
conclusions were based upon {\it Extreme UltraViolet Explorer (EUVE)} Deep Survey data (65-200 eV)
and {\it ROSAT} PSPC data (0.15-0.3 keV).
Subsequently, Bowyer, Berghoefer and Korpela (1999) reanalyzed the EUVE data and confirmed the existence of strong excess emission
in the central 15' of Coma. Arabadjis and Bregman (1999) reanalyzed the PSPC data, and reported that
the fitted HI column density in the center of the cluster was significantly smaller than the measured Galactic 
value, consistent with the earlier reports of soft X-ray excess emission.
Recently, the spatial distribution of the soft X-ray emission in the center of
the  Coma cluster 
was investigated by Bonamente et al. (2002) and Nevalainen et al. (2002).
In this  paper we present the analysis of a mosaic of {\it ROSAT} PSPC observations around the
Coma cluster, revealing a very diffuse soft X-ray halo extending to
considerably larger distances than reported in the previous studies.

The nature of the excess emission has been under active scrutiny.
The emission could originate from Inverse-Compton scattering of cosmic microwave background (CMB) photons
against a population of relativistic electrons in the intra-cluster medium (ICM), as advocated
by Hwang (1997), Sarazin and Lieu (1998), Ensslin and Biermann (1998) and Lieu et al. (1999). 
Alternatively, warm gas at T$\sim 10^6$ K could be responsible for the soft emission
(Lieu et al. 1996, Nevalainen et al. 2002).
Warm gas may reside inside the cluster, or in very diffuse filamentary structures
outside the cluster  projecting  onto it, as seen in
large scale hydrodynamic simulations (e.g., Cen and Ostriker 1999, Dav\'{e} et al. 2001, Cen et al 2001).
The warm gas scenario appears to be favored by the
current X-ray spectral analyses (e.g., Bonamente et
al. 2001a, Buote 2001), although a search for the UV OVI emission lines (1032-1039 \AA~in rest frame) 
has not yielded positive results (Dixon, Hurwitz and Ferguson 1996; Dixon et al. 2001). The non-detection of
emission lines can  be reconciled with the soft excess detection if the
gas has very low metal abundances, or alternatively if the gas exists in a temperature
range where OVI is not the predominant oxygen ion.
The spectral analysis of PSPC data reported in this paper indicates that the emission
is very likely thermal in nature.

The PSPC observations are described in section 2, and the distribution of the Galactic HI over the field of
view is presented in section 3.
Analysis of the PSPC spectra is given in section 4, followed by the interpretation in section 5  and 
conclusions in section 6. 

The redshift to the Coma cluster is $z$=0.023 (Struble and Rood 1999). Throughout this paper 
we assume a Hubble constant of $H_0=72$  km s$^{-1}$ Mpc$^{-1}$ (Freedman et al. 2001), 
and all quoted uncertainties are at the 68\% confidence level.

\section{The ROSAT PSPC data}

The {\it ROSAT} PSPC instrument has unique capabilities for the study of the low energy 
excess emission in clusters of galaxies.
Along with an effective area of $\sim 200$ cm$^{-2}$ at 0.25 keV, the
spectral response between 0.2 and 2 keV is well calibrated (Snowden et al. 1994).
Moreover, the large field of view ($\sim 1$ degree
radius), low detector background and  the availability of a large number of deep cluster observations
and of the {\it ROSAT} All-Sky Survey data (RASS),
render the PSPC a unique instrument to detect and investigate the large-scale
diffuse emission from galaxy clusters.

Coma is the nearest rich galaxy cluster, and the X-ray emission from its
hot ICM reaches an angular radius
of at least 1 degree (e.g., White et al. 1993, Briel et al. 2001). If the soft X-ray 
background level is estimated from the edge of the PSPC field of view
as is customarily done for more distant clusters,
the background can be significantly overestimated due to
the extended cluster emission (Bonamente et al. 2002). 
Here we show how  several off-center PSPC observations provide a reliable measurement of the background
and reveal a very extended halo 
of soft X-ray excess radiation, covering a
region  several megaparsecs in extent in the Coma cluster.

The RASS maps of the diffuse X-ray background (Snowden et al. 1997) are
suitable to measure the extent of the soft X-ray emission (R2 band, $\sim$ 0.15-0.3 keV) 
around the Coma cluster and to compare it with that 
of the higher-energy X-ray emission (R7 band, $\sim$ 1-2 keV). In Fig. 1 we show a radial
profile of the surface brightness of PSPC bands R2 and R7 centered on Coma: 
the higher-energy emission has reached a constant background value at a radius of $\sim$ 1.5 degrees, while the
soft X-ray emission persists out to a radius of $\sim$ 3 degrees. 
Thus, the RASS data indicate that the soft emission is more extended than the 1-2 keV emission, which 
originates primarily from the hot phase of the ICM.

The all-sky survey data is based on very short exposures ($\sim$ 700 sec); therefore, for detailed studies
of the soft X-ray emission in the Coma cluster, we use the four PSPC observations shown in Fig. 2.
Four additional deep PSPC observations (also shown in Fig. 2) are used to determine the local
background~\footnote{The ROSAT archival ID of the Coma pointings are RP8000005N00,RP8000006N00,RP8000009N00 and
RP8000013N00, the ID of the background fields are RP9002121A01,RP201514N00,
RP201471N00, and RP141917N00.}. We show in section 3 below that the distribution of $N_H$ is essentially constant within 5 degrees
of the cluster center. Therefore, the off-source PSPC fields located 2.5-4 degrees away from the cluster center
provide an accurate measurement of the soft X-ray background, and are also far enough away from the cluster
center to avoid being contaminated by cluster X-ray emission (Fig. 1).

\section{Galactic HI absorption in the direction of the Coma cluster}

Knowledge of the Galactic absorption is essential to determine the intrinsic X-ray emission from extragalactic
objects, particularly at energies $\leq$ 1 keV.
Throughout this paper, we use the absorption cross sections of Morrison and McCammon (1983) in our models.
These cross sections are in good agreement with the recent compilation
of Yan, Sadeghpour and Dalgarno (1998), as discussed by Arabadjis and Bregman (1999). At the resolution
of the PSPC, the He cross section in the most recent compilation by Wilms, Allen and McCray (2000) is
also indistinguishable from the values of Morrison and McCammon (1983), as discussed by Bonamente et al. (2002).

We use two methods to determine the distribution of  neutral hydrogen in the Coma cluster.
First, we use the radio measurements of Dickey and Lockman (1990) and of Hartmann and Burton (1997), which are
plotted in Fig. 3. The  radio measurements are in excellent agreement, and show the measured HI column density 
varying smoothly from 9 $\times 10^{19}$ cm$^{-2}$ to 11  $\times 10^{19}$ cm$^{-2}$
within a radius of 5 degrees from the center of the Coma cluster.
Second, we employ the far-infrared IRAS data, and use  the correlation between 100 $\mu$m IRAS flux and HI column density
(Boulanger and Perault 1988). The slope of the correlation is $1.2 \times 10^{20}$ cm$^{-2}$ / (MJy sr$^{-1}$),
and the offset is determined by fixing the central $N_H$ value to 9 $\times 10^{19}$ cm$^{-2}$, which is
well established from independent radio measurements of the center of the Coma cluster (Dickey and Lockman 1990,
Lieu et al. 1996, Hartmann and Burton 1997).
Fig. 3 shows that the radial variation of $N_H$ inferred from the IRAS data is in extremely good agreement with the
radio measurements. The data indicate that\\
 (a) the HI column density within the central 1.5 degree of the
cluster is constant (9$\pm$1 $\times 10^{19}$ cm$^{-2}$), and \\
(b) in the region where the off-source background fields are located (2.5-4 degrees from the
cluster center), the HI column density is between
9 $\times 10^{19}$ cm$^{-2}$ and 11 $\times 10^{19}$ cm$^{-2}$. An $N_H$ variation of this magnitude has a negligible effect
on the soft X-ray flux in the PSPC R2 band (cf. Fig. 5 in Snowden et al. 1998).

With the present IR and radio data we cannot  address the possibility of variations
in the HI distribution on  scales smaller that $\sim$ 10 arcmin.
On occasions for which a comparison with stellar Ly$\alpha$ and QSO X-ray spectra could be made,
the $N_H$ was found consistent with the wide-beam measurements 
to within $\sim 1 \times 10^{19}$ cm$^{-2}$ (Laor et al. 1994; Elvis, Lockman and
 Wilkes 1989; Dickey and Lockman 1990).

\section{Spectral analysis}

The 4 PSPC Coma observations were  divided into concentric annuli centered at
R.A.=12h59'48", Dec.=25$^o$57'0" (J2000), 
and the spectra were coadded to reduce the statistical errors.
The pointed PSPC
data were reduced according to the prescriptions of Snowden et al. (1994). The datasets
were corrected for detector gain fluctuations, and only events with Average Master Veto rate $\leq$ 170
c/s were considered, in order to discard periods of high particle background.
The PSPC rejection efficiency for particle background is 99.9 \% in the 0.2-2 keV energy range (Plucinsky
et al. 1993), and the background is therefore solely represented by the photonic component.
For each of the 4 off-source fields in Fig. 1, a spectrum was extracted after removal of 
point sources. The spectra were statistically consistent 
with one another within at most $\sim$ 10 \% 
point-to-point fluctuations. The off-source spectra were  therefore coadded,
and a 10\% systematic uncertainty in the background was included 
in the error analysis.
 Further details of the PSPC data analysis
can be found in Bonamente et al. (2002).

\subsection{Single temperature fits}
Initially, we fit the spectrum of each annulus in XSPEC, using a single-temperature MEKAL plasma model 
(Mewe, Gronenschild and van den Oord 1985;
 Mewe, Lemen and van den Oord 1986; Kaastra 1992)
and the WABS Galactic absorption model (Morrison and McCammon 1983).
The results of the single temperature fit are given in Table 1, and are shown for one of the annuli in Fig. 4. If the 
neutral hydrogen column density is fixed at the Galactic value (9 $\times 10^{19}$ cm$^{-2}$, see section 3),
the fits are statistically unacceptable (reduced $\chi^2$ ranging from 3.1 to 9.7).
Allowing the neutral hydrogen column density to vary results in an
unrealistically low $N_H$ for all of the annuli, and also produces statistically 
unacceptable fits (reduced $\chi^2$ ranging from 2.5 to 4.2).
We conclude that a single temperature plasma model does not adequately describe the spectral data,
particularly at energies below 1 keV (Fig. 4).
Therefore, in the analysis that follows, we fit only the high energy portion of the spectrum (1-2 keV)
with a single temperature plasma model, and introduce an additional model component to account for the 
low energy emission.

\subsection{Modelling the hot ICM}
To fit the high energy portion of the spectrum, we apply a MEKAL model to the data between 1 and 2 keV, and a 
photoelectric absorption model with $N_H=9 \times 10^{19}$ cm$^{-2}$. The metal abundance is fixed at 0.25 solar
for the central 20 arcmin region (Arnaud et al. 2001), and at 0.2 solar in the outer regions. 
The spectra are also subdivided into quadrants, in order to obtain 
a more accurate temperature for each region of the cluster.
The results of the `hot ICM' fit are given in Table 2, and are consistent with the results
previously derived from the PSPC data by Briel and Henry (1997),
and with  recent XMM measurements (Arnaud et al. 2001).
In addition, the temperature found at large radii is in agreement with the composite
cluster temperature profile of De Grandi and Molendi (2002).

\subsection{Soft excess emission}
The measured  fluxes in the soft  X-ray band can now be compared with the
hot ICM model predictions in the 0.2-1 keV band. The results are shown in Table 2
and Fig. 5. The error bars reflect the uncertainty in the hot ICM temperature (Table 2)
and the uncertainty in the Galactic HI column density 
($N_H=9\pm1 \times 10^{19}$ cm$^{-2}$).
The soft excess component  is detected with high statistical significance 
throughout the 90' radius of the pointed PSPC data, which corresponds to
a radial distance of 2.6  Mpc. The soft excess emission
(Fig. 5, left panel) is much more extended than that of the hot ICM (Fig. 5, right panel),
in agreement with the conclusions drawn from the all-sky survey data (Fig. 1).

\subsection{Low energy non-thermal component}
Having established the hot ICM temperature for each quadrant (Table 2), we now consider additional components
in the spectral analysis. First, we add a power law non-thermal component, which predominantly 
contributes to the low energy region of the spectrum (Sarazin and Lieu 1998). The neutral
hydrogen column density was fixed at the Galactic value ($N_H=9 \times 10^{19}$ cm$^{-2}$).
The results of fitting the hot ICM plus power law models to the annular regions are shown in Table 3.
The reduced $\chi^2$ values are poor: the average $\chi^2_{red}$ is 1.48, and the worst case value is 1.89;
we conclude that the combination of a low energy power law component and the hot ICM thermal model does not adequately
describe the PSPC spectral data.

\subsection{Low energy thermal component}
Finally, we consider a model consisting of a `hot ICM' thermal component (section 4.2)
and an additional low-temperature thermal component. As before, the neutral hydrogen column density 
was fixed at the Galactic value (see section 3). 
The results of fitting the hot ICM plus warm thermal models are shown in Table 3. The reduced $\chi^2$ values are
significantly improved relative to the previous case:
the average reduced $\chi^2$ is 1.24 and the worst case value is 1.45. In every region the 
fit obtained with a warm thermal component was superior to the fit using a non-thermal component, as
indicated by inspection of the $\chi^2_{red}$ values and by an F-test (Bevington 1969) on the
two $\chi^2$ distributions (Table 3).

\section{Interpretation}

The spectral analysis of section 4 indicates that the excess emission can be explained
as thermal radiation from diffuse warm gas. The non-thermal model
appears viable only in a few quadrants, and will not be further considered in this paper.

\subsection{A warm phase of the ICM}
If the soft excess emission originates from a warm phase of the intra-cluster medium,
 the ratio of the emission integral of the hot ICM (Table 2) and  the 
emission integral of the warm gas (Table 3) 
can be used to measure the relative mass of the two phases.
The emission integral is defined  as
\begin{equation}
I=\int n^2 dV \;,
\end{equation}
where $n$ is the gas density and $dV$ is the volume of emitting region (Sarazin 1988).
The emission integral is readily measured by fitting the X-ray spectrum.~\footnote{See 
description of XSPEC MEKAL model at http://heasarc.gsfc.nasa.gov/docs/xanadu/xspec/manual/manual.html.}

The emission integral of each quadrant determines the average density of the  gas
in that region, once
the volume of the emitting region is specified (Eq. 1). Assuming that each 
quadrant corresponds to a sector of a spherical shell, the density
in each sector can be calculated. 
The density of the warm gas ranges from $\sim$ 9$\times 10^{-4}$ cm$^{-3}$ 
to $\sim 8 \times 10^{-5}$ cm$^{-3}$, and the  density of the hot gas
varies from $1.5 \times 10^{-3}$ cm$^{-3}$ 
to $6 \times 10^{-5}$ cm$^{-3}$.

We assume that both the warm gas and the hot gas are distributed in spherical shells of constant density.
 Since the emission integral is proportional to $n^2 dV$ and the mass is proportional
to $n dV$, the ratio of the warm-to-hot gas mass is
\begin{equation}
\frac{M_{warm}}{M_{hot}}=\frac{\int n_{warm} dV}{\int n_{hot} dV} =  \frac{\int dI_{warm}/n_{warm}}{\int dI_{hot}/n_{hot}} 
\end{equation} 
We evaluate Eq. 2 by summing  the values of $I_{hot}/n_{hot}$ and
$I_{warm}/n_{warm}$ for all regions (Tables 2 and 3),
and conclude that  $M_{warm}/M_{hot}$=0.75 within a radius of 2.6 Mpc.

\subsection{Warm filaments around the Coma cluster}
It is also possible that the warm gas is distributed in extended low-density
filaments rather then being concentrated near the cluster center like the hot ICM.
Recent large-scale hydrodynamic  simulations 
(e.g., Cen et al. 2001, Dav\'{e} et al. 2001, Cen and Ostriker 1999) indicate that
this is the case, and that 30-40 \% of the present epoch's baryons reside in these
filamentary structures. Typical filaments feature a temperature of T$\sim10^5-10^7$ K,
consistent with our results in Table 3, and density of $\sim 10^{-5}-10^{-4}$ cm$^{-3}$ 
(overdensity of $\delta \sim 30-300$, Cen et al. 2001).

The ratio of mass in warm filaments to mass in the hot ICM is
\begin{equation}
\frac{M_{fil}}{M_{hot}}= \frac{\int n_{fil} dV_{fil}}{\int n_{hot} dV}= \frac{\int dI_{warm}/n_{fil}}{\int dI_{hot}/n_{hot}}.
\end{equation}

Assuming a filament density of $n_{fil}=10^{-4}$ cm$^{-3}$, Eq. (3) yields the conclusion that $M_{fil}/M_{hot}$= 3 within
a radius of 2.6 Mpc; the ratio will be even larger if the filaments are less dense.
The warm gas is therefore  more massive than the hot ICM if it is distributed in low-density
filaments. More detailed mass estimates require precise knowledge of the filaments spatial distribution.

\section{Conclusions}

The analysis of deep PSPC data of the Coma cluster reveals a large-scale
halo of soft excess radiation, considerably more extended than previously thought.
The PSPC data indicate that the excess emission is due to warm gas
at T$\sim 10^6$ K, which may exist either as
a second phase of the intra-cluster medium,
or in diffuse filaments outside the cluster. Evidence in favor of the latter scenario
is provided by the fact that the spatial extent of the
soft excess emission is significantly greater than  that of the hot ICM.

The total mass of the Coma cluster within 14 Mpc is 1.6$\pm0.4 \times 10^{15} M_{\bigodot}$
(Geller, Diaferio and Kurtz).
The mass of the  hot ICM is
$\sim 4.3 \times 10^{14} M_{\bigodot}$ within 2.6 Mpc (Mohr, Mathiesen and Evrard 1999).
The present detection of soft excess emission out to a distance of 2.6 Mpc from the cluster's
center implies that the warm gas has a mass of at least $ 3 \times 10^{14} M_{\bigodot}$,
or considerably  larger if the gas is in very low density filaments.
The PSPC data presented in this paper 
therefore lends observational support to the current theories of large-scale formation and evolution (e.g,
Cen and Ostriker 1999),
which predict that a large fraction of the current epoch's baryons are in
a diffuse warm phase of the intergalactic medium.

\begin{deluxetable}{lccccccc}
\tabletypesize{\small}
\tablecaption{Best-fit single temperature models}
\tablehead{ & \multicolumn{3}{c}{0.2-2 keV fit, Galactic $N_H$} & \multicolumn{4}{c}{0.2-2 keV fit, free $N_H$}\\
      & \multicolumn{3}{c}{\hrulefill} & \multicolumn{4}{c}{\hrulefill} \\
Region & kT & A & $\chi^2_{r}$/d.o.f & $N_H$ & kT & A & $\chi^2_{r}$/d.o.f\\
          (arcmin)  & (keV) & &                    & ($10^{19}$ cm$^{-2})$ & (keV) & & }
\startdata
 0-20 & 3.9 $\pm0.15$ & 0.25 & 9.74/181 & 6.15 $\pm0.1$ & 6.5 $\pm^{0.4}_{0.3}$ & 0.3 & 2.8/180 \\
20-40 & 2.4  $\pm0.1$ & 0.2 & 7.4/181 & 2 $\pm{0.3}$ & 3 $\pm^{0.3}_{0.1}$ & 0.2 & 3.7/180 \\
40-55 & 2.3 $\pm^{0.1}_{0.1}$ & 0.2 & 6.2/175 & $\leq$ 0.01 & 2.3 $\pm 0.1$ & 0.2 & 4.2/174 \\
50-70 & 1.8 $\pm^{0.3}_{0.1}$ & 0.2 & 3.5/151 &  $\leq$ 0.02 & 1.85 $\pm^{0.4}_{0.1}$ & 0.2 & 2.9/150 \\
70-90 & 1.9 $\pm^{0.3}_{0.1}$ & 0.2 & 3.1/139 & $\leq$ 0.02 & 2$\pm^{0.5}_{0.3}$ & 0.2 & 2.5/138 \\
\enddata
\end{deluxetable}

\begin{deluxetable}{rccccc}
\tabletypesize{\small}
\tablecaption{Best-fit hot ICM model and soft X-ray fluxes}
\tablehead{ & \multicolumn{3}{c}{Hot ICM} & \multicolumn{2}{c}{soft X-ray fluxes$^{(*)}$} \\
    & \multicolumn{3}{c}{\hrulefill} & \multicolumn{2}{c}{\hrulefill} \\
Region & kT   & $I^{(**)}$ & $\chi^2_{r}$/dof & measured flux &  hot ICM prediction$^{(***)}$ \\
 (arcmin) & (keV) &  & & (c s$^{-1}$) & (c s$^{-1}$) }
\startdata
0-20 ALL& 7 $\pm{0.25}$  & 23.6$\pm$0.1 & 1.42/100 &  8.85$\pm$0.016 & 8.0 $\pm^{0.05}_{0.075}\pm^{0.22}_{0.23}$\\
 NE & 5.8 $\pm{0.4}$   & 4.9$\pm$0.1& 0.82/100 & 2.03$\pm$0.007 & 1.81$\pm^{0.02}_{0.02}\pm^{0.06}_{0.04}$\\
 NW & 6.6 $\pm{0.6}$   &  5.0$\pm$0.05  &0.89/98 & 1.86$\pm$0.007& 1.7$\pm^{0.035}_{0.025}\pm^{0.05}_{0.04}$\\
 SE & 6.8 $\pm{0.5}$   & 6.6$\pm$0.05&0.82/98 & 2.42$\pm$0.008 & 2.18$\pm^{0.03}_{0.03}\pm^{0.06}_{0.06}$\\
 SW & 13.6 $\pm^{2.9}_{2}$ & 7.7$\pm$0.05 & 1.01/98 & 2.27$\pm$0.007 & 2.0$\pm^{0.08}_{0.08}\pm^{0.06}_{0.05}$\\
\hline
20-40 ALL& 4.2 $\pm{0.3}$  &6.1$\pm$0.05 & 0.98/100 & 2.65$\pm$0.02 & 1.88$\pm^{0.02}_{0.02}\pm^{0.05}_{0.05}$ \\
 NE & 2.6 $\pm{0.3}$  & 0.85$\pm$0.015&  0.69/98 & 0.53$\pm$0.006 & 0.345$\pm^{0.005}_{0.005}\pm^{0.01}_{0.01}$\\
 NW & 4.3 $\pm{0.7}$  & 1.65$\pm$0.05& 0.76/98 & 0.685$\pm$0.007 & 0.50$\pm^{0.011}_{0.013}\pm^{0.015}_{0.015}$\\
 SE & 8 $\pm^{3.2}_{2}$ & 1.55$\pm$0.03& 0.57/98 & 0.632$\pm$0.007  & 0.42$\pm^{0.025}_{0.027}\pm^{0.013}_{0.011}$ \\
SW & 7.8$\pm^{2}_{1.4}$ &  3.3$\pm$0.1 &0.89/98& 0.935$\pm$0.007 & 0.714$\pm^{0.03}_{0.04}\pm^{0.02}_{0.02}$\\
\hline
40-55 ALL& 5.3 $\pm^{2.2}_{1.3}$ & 2.9$\pm$0.1 & 0.89/95 & 1.8$\pm$0.038 &0.65$\pm^{0.03}_{0.035}\pm^{0.015}_{0.02}$  \\
 NE & 3.1$\pm^{3}_{1}$ & 0.33$\pm$0.03 &0.69/50 &  0.36$\pm$0.01 & 0.095$\pm^{0.005}_{0.01}\pm^{0.003}_{0.002}$\\
 NW & 8.0$\pm^{7}_{3}$ & 0.42$\pm$0.03&0.93/54 &  0.68$\pm$0.015 & 0.18$\pm^{0.013}_{0.027}\pm^{0.005}_{0.004}$\\
 SE & 3.6 $\pm^{8}_{1.3}$  & 0.34$\pm$0.03&   0.96/48 & 0.38$\pm$0.01& 0.091$\pm^{0.006}_{0.026}\pm^{0.004}_{0.025}$ \\
 SW & 3.5 $\pm{0.7}$  & 1.6$\pm$0.05& 1.0/78 &  0.67$\pm$0.01 & 0.38$\pm^{0.015}_{0.01}\pm^{0.015}_{0.01}$\\
\hline
55-70 ALL & 2.6 $\pm^{0.9}_{0.5}$ & 0.98$\pm$0.05 & 0.97/136 & 0.85$\pm$0.024 & 0.3$\pm^{0.006}_{0.007}\pm^{0.009}_{0.008}$\\
 NE & 3.5 $\pm^{4.5}_{1.3}$  & 0.5$\pm$0.04& 1.02/70 &  0.5$\pm$0.02& 0.137$\pm^{0.012}_{0.019}\pm^{0.005}_{0.004}$\\
 SW & 2.3 $\pm^{0.8}_{0.5}$  & 0.48$\pm$0.04 & 0.93/65 &  0.35$\pm$0.014 & 0.16$\pm^{0.005}_{0.004}\pm^{0.006}_{0.005}$\\
\hline
70-90 SW& 2.9 $\pm^{2.5}_{0.8}$  & 0.42$\pm$0.03& 1.21/59 & 0.335$\pm$0.014 & 0.121$\pm^{0.005}_{0.01}\pm^{0.005}_{0.004}$ \\
\enddata
\tablecomments{
(*) Soft X-ray fluxes are in the 0.2-1 keV band.\\
(**) $I$ is the best-fit emission integral in the units of $(10^{-16}/ 4 \pi ((1+z) D)^2)$, where $D$ is
the distance to the source (in cm) and  $z$ is the redshift.\\
(***) The two error brackets account respectively for the uncertainty in the hot ICM temperature and the uncertainty in the
Galactic HI column density ($\Delta(N_H)=\times 10^{19}$ cm$^{-2}$).}
\end{deluxetable}

\begin{deluxetable}{rccccccc}
\tabletypesize{\small}
\tablecaption{Spectral models}
\tablehead{  &   \multicolumn{2}{c}{Non-thermal component} & \multicolumn{4}{c}{ Low energy thermal component} \\
     &   \multicolumn{2}{c}{\hrulefill} & \multicolumn{4}{c}{\hrulefill} & \\
Region &  $\alpha$ & $\chi^2_{r}$/d.o.f  &  kT   & A & $I$ & $\chi^2_{r}$/d.o.f &  F-test \\
 (amin)  &  & & (keV) &   &  & &  probability~$^{(*)}$ }
\startdata
0-20 NE & 2.65$\pm{0.05}$ &1.42/180 &   0.29$\pm0.05$  & 0.02$\pm0.015$ & 1.3$\pm$0.25 & 1.36/179 & 61.3\%\\
NW & 1.42$\pm0.01$ & 1.54/180 &  0.25$\pm0.05$  & 0.05$\pm0.03$ & 2.0$\pm{1.0}$ & 1.18/179 & 96.2\%  \\
 SE & 1.51$\pm{0.02}$ & 1.52/180&  0.12$\pm0.015$ & 0.07$\pm0.03$ & 1.5$\pm$0.5 &1.36/179 &77.1\%  \\
 SW &1.40$\pm{0.005}$ & 1.45/180&  0.093$\pm{0.01}$ & 0.1 & 0.93$\pm$0.06 &  1.41/180 & 57.4\%\\
\hline
20-40 NE & 1.93$\pm{0.07}$ & 1.45/180 &  0.22$\pm{0.015}$ & 0.06$\pm^{0.05}_{0.02}$ & 0.7$\pm$0.08 &   1.25/179 & 83.9\%  \\
 NW &1.73$\pm{0.07}$ & 1.59/180 &  0.21$\pm{0.015}$ & 0.07$\pm^{0.06}_{0.03}$ & 0.8$\pm$0.09 &  1.45/179 &73.1\% \\
 SE &  1.70$\pm^{0.07}_{0.03}$ & 1.3/180&  0.22$\pm{0.015}$ & 0.06$\pm^{0.04}_{0.02}$ &  0.95$\pm$0.06 &   1.1/179 & 86.8\% \\
 SW &2.30$\pm{0.15}$ & 1.43/180  & 0.23$\pm 0.08$ & 0.04$\pm 0.025$ & 1.45$\pm$0.1  & 1.33/179  & 68.6\%\\
\hline
40-55 NE & 2.37$\pm{0.05}$ & 1.64/131 &  0.21$\pm{0.015}$ & 0.1$\pm^{0.1}_{0.04}$ & 1.15$\pm$0.2 &   1.1/130 &98.8\% \\
 NW &2.37$\pm^{0.07}_{0.04}$ & 1.15/134 & 0.21$\pm{0.025}$ & 0.03$\pm^{0.035}_{0.02}$ & 1.7 $\pm$0.15 &  1.06/133  &68.1\% \\
 SE & 2.44$\pm{0.06}$ & 1.59/130&  0.215$\pm{0.015}$ & 0.05$\pm^{0.05}_{0.015}$ & 1.7 $\pm$ 0.2 & 1.24/129 & 92.1\%  \\
 SW & 1.91$\pm{0.05}$ & 1.15/160&  0.215$\pm{0.025}$ & 0.03$\pm^{0.035}_{0.015}$ & 2.0$\pm$0.2 & 1.15/159 & 50.0\% \\
\hline
55-70 NE   & 2.35$\pm 0.06$ & 1.69/150& 0.20$\pm0.02$ & 0.18$\pm^{0.17}_{0.08}$ & 1.2$\pm$0.45 & 1.08/149  &  99.7\% \\
 SW & 2.21$\pm{0.15}$ & 1.40/144 & 0.23$\pm{0.025}$ & 0.12$\pm^{0.07}_{0.05}$ & 0.7$\pm$0.15 &1.10 /143  &  92.5\% \\
\hline
70-90 SW & 2.2$\pm^{0.05}_{0.1}$ & 1.89/140&  0.22$\pm{0.02}$ & 0.15$\pm^{0.1}_{0.04}$ & 0.7$\pm$0.2 &1.41/139  & 95.8\% \\
\enddata
\tablecomments{(*) The F-test describes the likelihood that the thermal model consititutes an improvement
over the non-thermal model (Bevington 1969). }
\end{deluxetable}

\begin{figure}
       \epsscale{0.8}
       \plotone{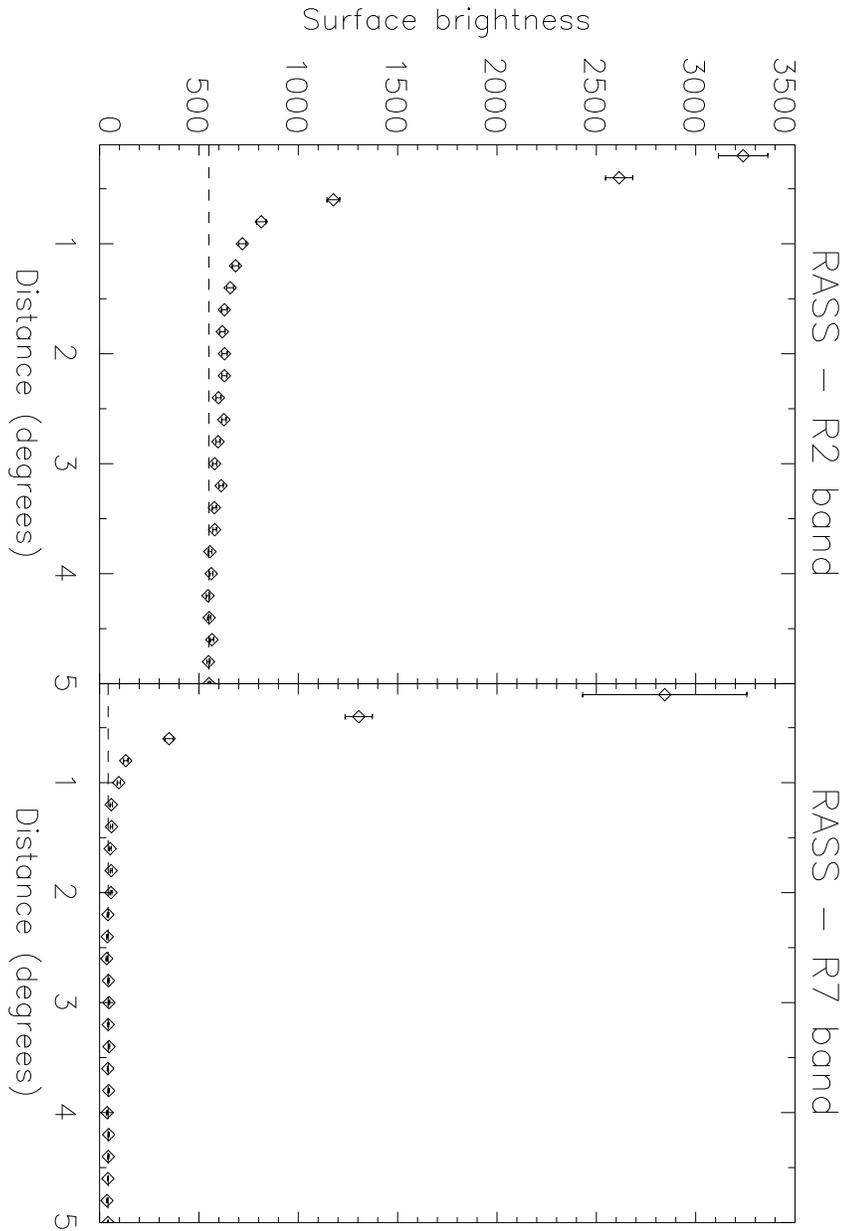}
       \caption{Radial profiles 
of soft X-ray emission (R2 band, 0.15-0.3 keV) and higher energy X-ray emission
(R7 band, 1-2 keV) in the Coma cluster. Surface brightness units are
$10^{-6}$ counts s$^{-1}$ arcmin$^{-2}$ pixel$^{-1}$ (Snowden et al. 1998)}
       \end{figure}

\begin{figure}
       \epsscale{0.8}
       \plotone{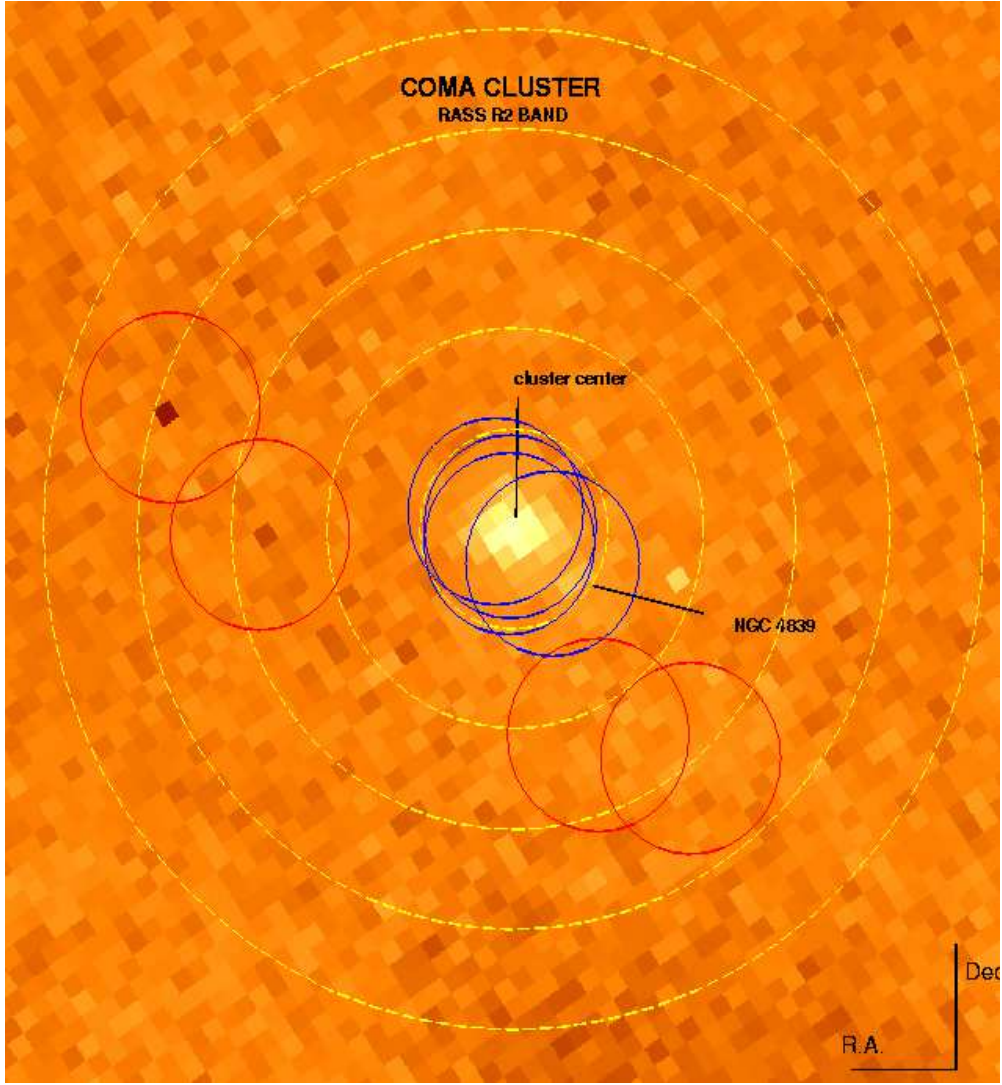}
       \caption{Location of {\it ROSAT} PSPC observations, overlaid on 
a RASS R2 band (0.15-0.3 keV) image of the diffuse emission of the Coma region. Dashed circles
indicate distance from the cluster's center in intervals of 1 degree, blue circles represent the
position of the pointed PSPC observations of Coma cluster, red circles are background observations.}
       \end{figure}

\begin{figure}
       \epsscale{0.8}
       \plotone{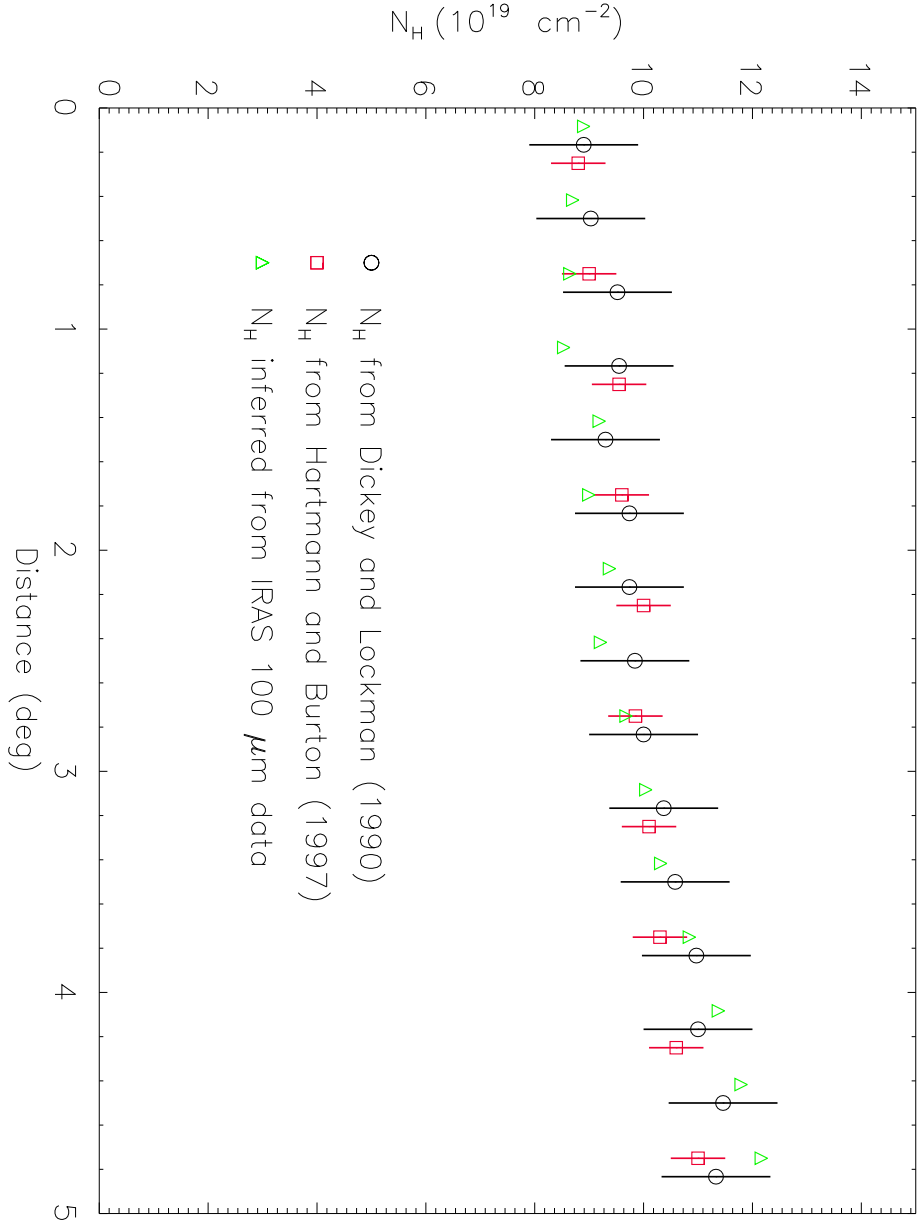}
       \caption{The Galactic $N_H$ distribution around the Coma cluster from the Dickey and Lockman (1990) radio
survey data, the Hartmann and Burton (1997) radio data, and as inferred from IRAS 100 $\mu$m data. Errors
in the IRAS measurements are estimated to be less than 2 $\times 10^{19}$ cm$^{-2}$ (Boulanger and Perault 1988).
}
       \end{figure}

\begin{figure}
       \epsscale{0.8}
       \plotone{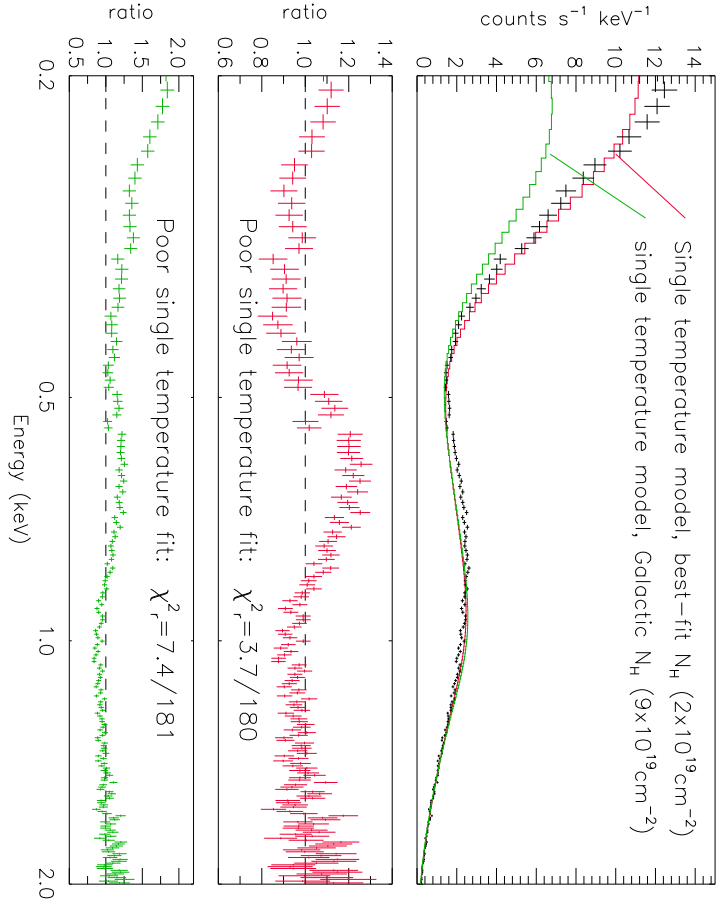}
       \caption{{\it ROSAT} PSPC spectrum of the 20-40' region around the center of the
Coma cluster, fitted to a single temperature model with variable $N_H$ 
(red), and with Galactic $N_H$ (green).}
       \end{figure}

\begin{figure}
       \epsscale{0.8}
       \plotone{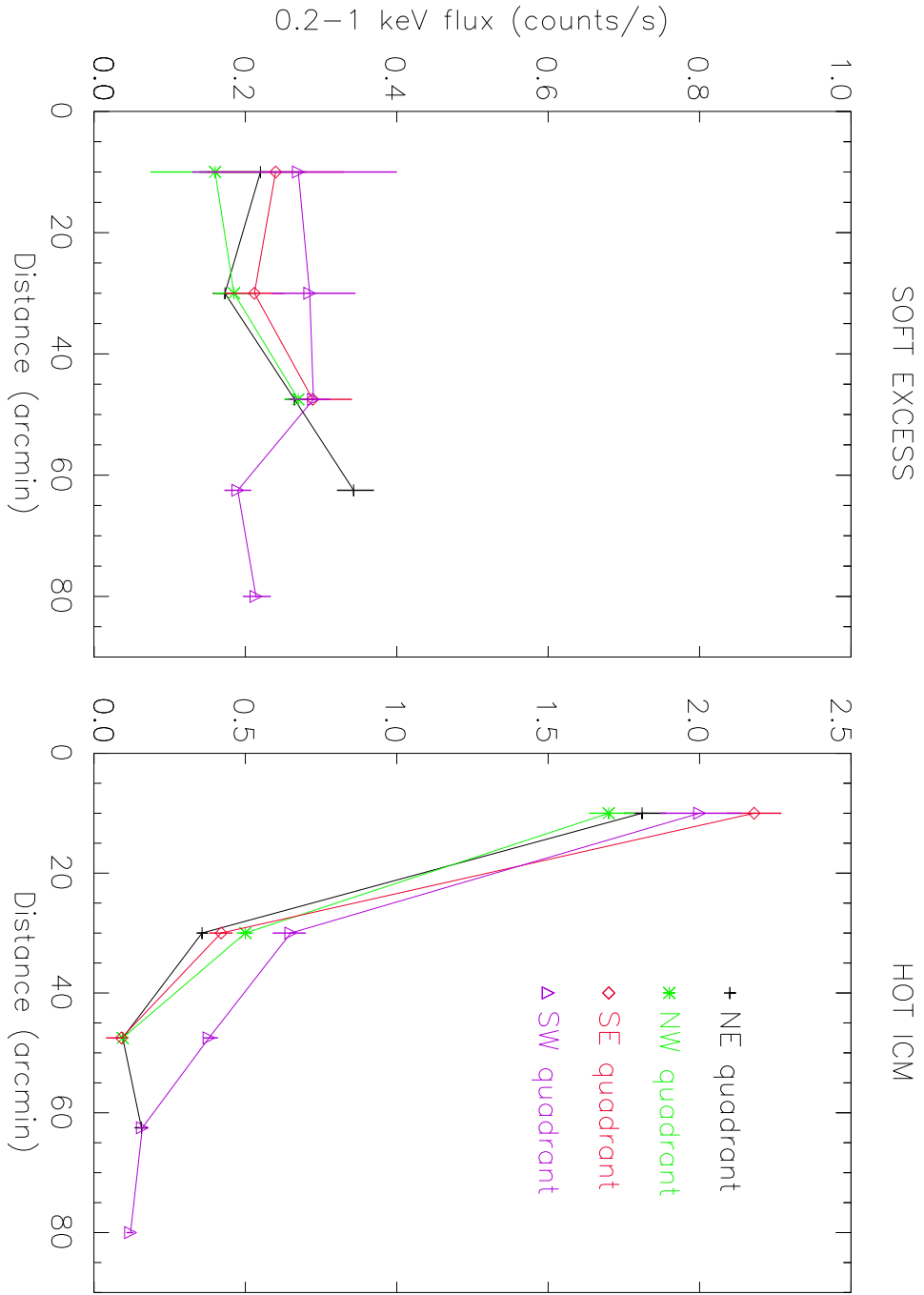}
       \caption{The radial distribution of the soft excess emission and of emission from the hot ICM}
       \end{figure}

\begin{figure}
       \epsscale{0.8}
       \plotone{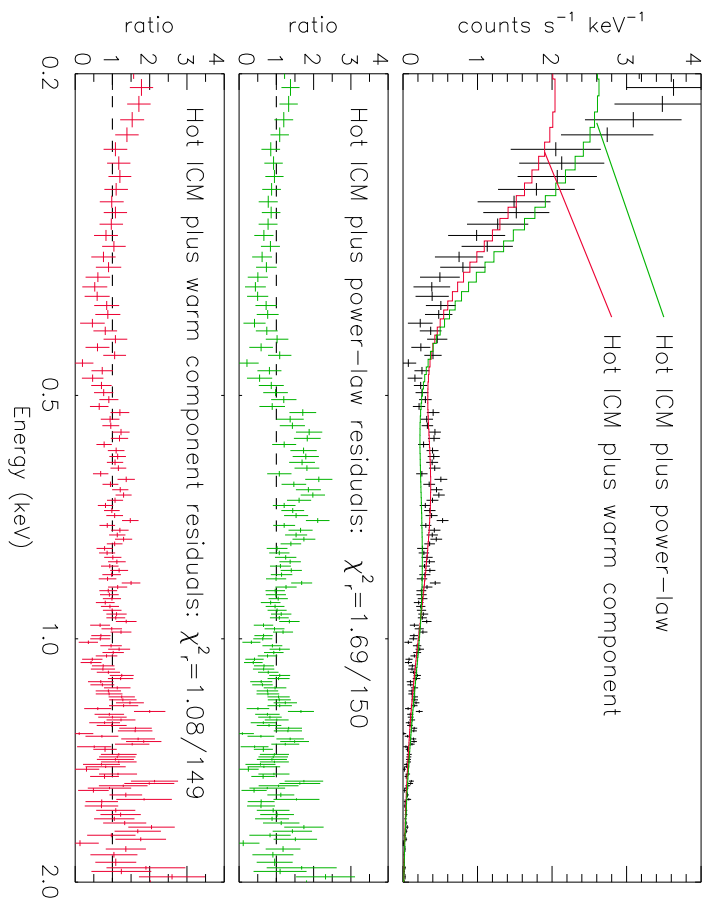}
       \caption{ {\it ROSAT} PSPC spectrum of the 50-70' north-eastern quadrant. In green is the hot ICM
 plus power law model (section 4.4), in red the hot ICM model plus a low-energy thermal component (section 4.5).}
       \end{figure}

\end{document}